\documentclass[10pt,twocolumn,english,aps,prl,nofootinbib]{revtex4}
\usepackage[utf8]{inputenc}
\usepackage{graphicx}
\usepackage{amsmath}
\usepackage{amssymb}
\usepackage{amsfonts}
\usepackage{lscape}
\usepackage{hyperref}
\usepackage{url}
\usepackage{epsfig}
\usepackage{color}
\usepackage{bm}
\usepackage{tabularx}
\newcommand{\be}{\begin{equation}}
\newcommand{\ee}{\end{equation}}
\newcommand{\ba}{\begin{eqnarray}}
\newcommand{\ea}{\end{eqnarray}}
\newcommand{\nn}{\nonumber}
\newcommand{\Msun}{M_\odot}

\renewcommand{\[}{\begin{equation}}
\renewcommand{\]}{\end{equation}}

\makeatother

\begin{document}

\title{Gravitational wave bursts from Primordial Black Hole hyperbolic encounters}

\author{Juan Garc\'ia-Bellido$^{(a,b)}$}
\email{juan.garciabellido@cern.ch}

\author{Savvas Nesseris$^{(a)}$}
\email{savvas.nesseris@csic.es}

\affiliation{$^{(a)}$Instituto de F\'isica Te\'orica UAM-CSIC, Universidad Auton\'oma de Madrid,
Cantoblanco, 28049 Madrid, Spain\\
$^{(b)}$CERN, Theoretical Physics Department, 1211 Geneva, Switzerland}

\date{\today}

\begin{abstract}
We propose that Gravitational Wave (GW) bursts with millisecond durations can be explained by the GW emission from the hyperbolic encounters of Primordial Black Holes in dense clusters. These bursts are single events, with the bulk of the released energy happening during the closest approach, and emitted in frequencies within the AdvLIGO sensitivity range. We provide expressions for the shape of the GW emission in terms of the peak frequency and amplitude, and estimate the rates of these events for a variety of mass and velocity configurations. We study the regions of parameter space that will allow detection by both AdvLIGO and, in the future, LISA. We find for realistic configurations, with total mass $M\sim60\,M_\odot$, relative velocities $v\sim 0.01\,c$, and impact parameters $b\sim10^{-3}$ AU, for AdvLIGO an expected event rate is ${\cal O}(10)$ events/yr/Gpc$^3$ with millisecond durations. For LISA, the typical duration is in the range of minutes to hours and the event-rate is ${\cal O}(10^3)$ events/yr/Gpc$^3$ for both $10^3\,M_\odot$ IMBH and $10^6\,M_\odot$ SMBH encounters. We also study the distribution functions of eccentricities, peak frequencies and characteristic timescales that can be expected for a population of scattering PBH with a log-normal distribution in masses, different relative velocities and a flat prior on the impact parameter.
\end{abstract}
\maketitle

\section{Introduction}

Advanced LIGO has opened a new era of Gravitational Wave Astronomy, with the detection of at least three very massive Black Hole (BH) merger events~\cite{Abbott:2016blz,Abbott:2016nmj,Abbott:2017vtc}, and probably a fourth one~\cite{TheLIGOScientific:2016pea}, in a few months of running. The signal corresponds to the inspiralling of two BH of several tens of solar masses in almost circular orbits, and the emission of GW leading to the final merger is in agreement,  within experimental errors, with the predictions of General Relativity (GR). These massive BH binaries were rather unexpected, see however~\cite{Belczynski:2009xy}, suggesting a new population of very massive BH. This led to the speculation that AdvLIGO could have detected Primordial Black Holes (PBH) contributing to a significant fraction of Cold Dark Matter (CDM)~\cite{Bird:2016dcv,Clesse:2016vqa,Sasaki:2016jop}, thus providing a natural explanation for its nature without resorting to exotic particles or modifications of gravity. Furthermore, these PBH could also provide the seeds for the Supermassive Black Holes (SMBH) found in the centers of the galaxies, as well as explaining the missing satellite and too-big-to-fail problems of CDM, thus solving several key problems in cosmology and galaxy formation in a unique and unified framework~\cite{Garcia-Bellido:2017fdg}.

In the scenario of {\em clustered} PBH of Ref.~\cite{Clesse:2015wea}, it is expected that a large fraction of BH encounters will not end up producing bounded systems, which would then inspiral, but rather produce a single scattering event, via a hyperbolic encounter. This could happen, e.g. if the relative velocity or relative distance of the two PBH is high enough that capture is not possible. The emission of GW in parabolic and hyperbolic encounters of compact bodies has been studied in the past in Refs.~\cite{OLeary:2008myb}, and~\cite{Capozziello:2008ra,DeVittori:2012da}, respectively. These events generate {\em bursts} of gravitational waves, which can be sufficiently bright to be detected at distances up to several Gpc. For clustered PBH, the waveform and characteristic parameters of the GW emission in hyperbolic encounters are different to those of the inspiralling binaries, and both provide complementary information that can be used to determine the evolved mass distribution of PBH, as a function of redshift, as well as their spatial distribution.

Hyperbolic encounters are single scattering events where the majority of the energy is released near the point of closest approach, and have a characteristic peak frequency which is a function of only three variables, the impact parameter $b$, the eccentricity $e$ and the total mass of the system $M$. Furthermore, the duration of such events is of the order of a few milliseconds to several hours, depending on those parameters. The case of inspiralling and merging PBH has been studied extensively, see e.g. Refs.~\cite{Clesse:2016vqa,Clesse:2016ajp}, and estimated to produce a few tens of events/year/Gpc$^3$ in the range of $M_{\rm PBH} \sim {\cal O}(10-100)\,\Msun$. In this letter we will show that, within the parameter space of the clustered PBH scenario~\cite{Clesse:2015wea,Clesse:2016vqa}, we expect a similar but somewhat lower rate of GW burst events in the millisecond range.

For a detector like AdvLIGO, such events will look like bursts with a characteristic frequency at peak strain amplitude. In fact, AdvLIGO has reported a few events of this type, which were attributed to accidental noise in the detectors~\cite{Abbott:2016ezn}. However, events from hyperbolic encounters of PBH produce shapes that are rather similar to the ``tear drop glitch" described in Ref.~\cite{Powell:2016rkl}. It is thus worth exploring the possibility that some of those events are actually PBH hyperbolic encounters. Their time-frequency profiles, discussed in this letter, could help the analysis of the Adv\-LIGO bursts and glitches. Moreover, if indeed these glitches arise from hyperbolic encounters of BH, they could be used to obtain valuable information about the PBH mass, velocity and spatial distribution.

\begin{figure}[!h]
\centering
\includegraphics[width = 0.48\textwidth]{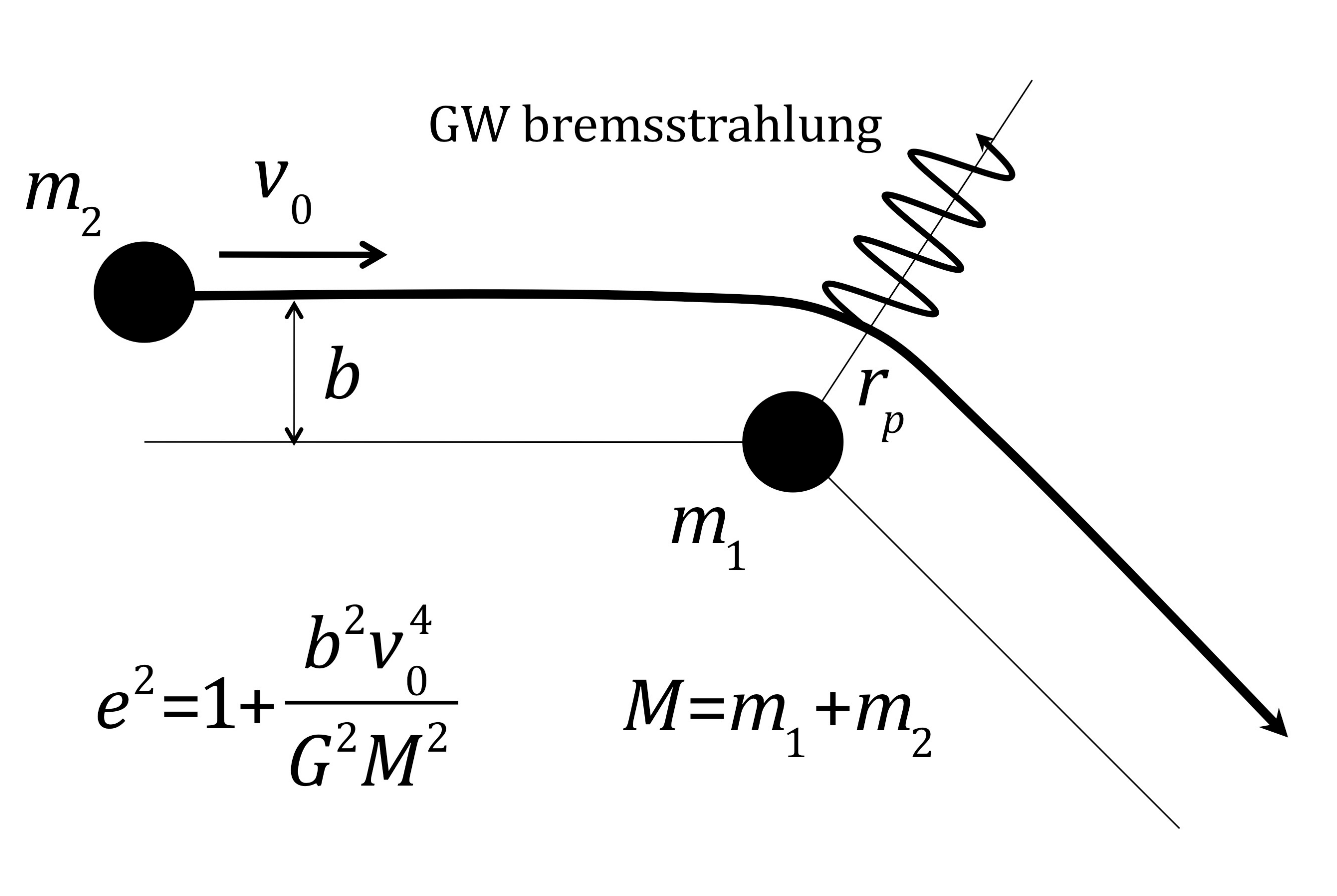}
\caption{The scattering of one BH of mass $m_2$ on another of mass $m_1$ induces the emission of gravitational waves which is maximal at the point of closest approach, $r_p$.}
\vspace*{-1mm}
\label{fig:hyperbolic}
\end{figure}

\section{Hyperbolic encounters of PBH}

\begin{figure}[!h]
\centering
\hspace*{-4mm}
\includegraphics[width = 0.42\textwidth]{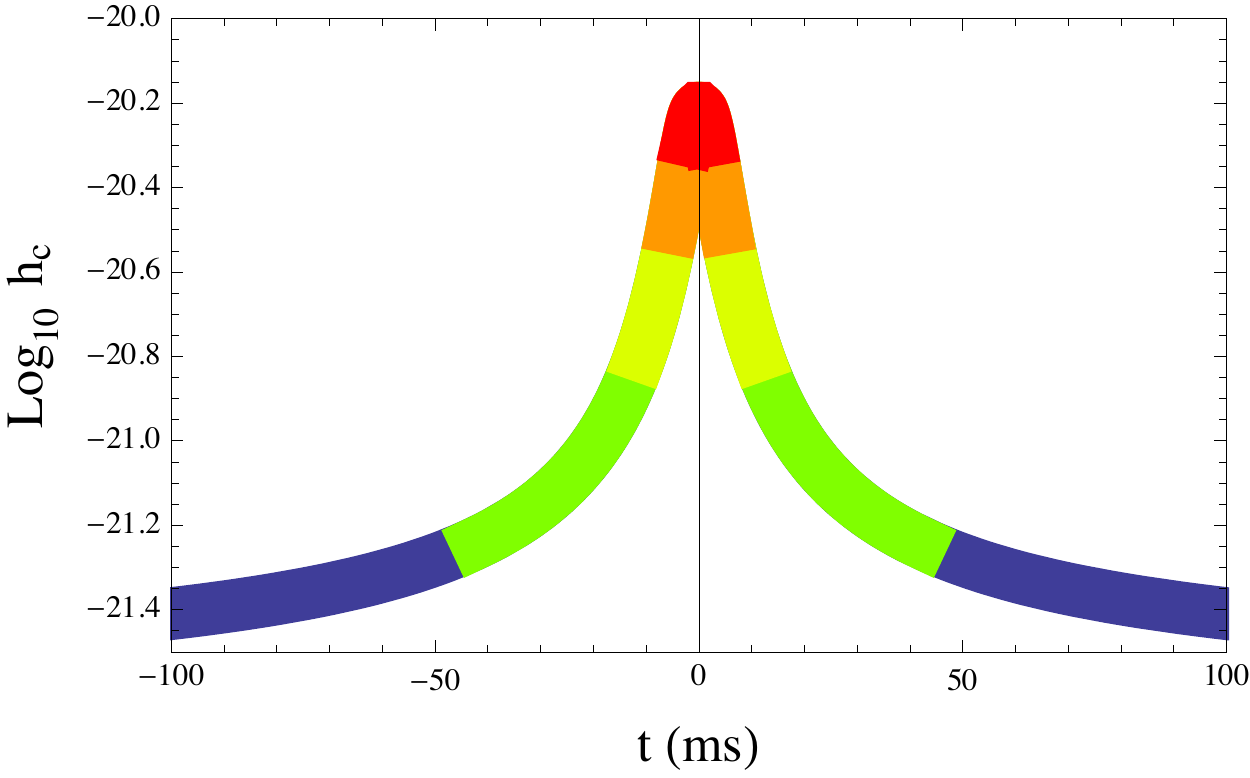}
\includegraphics[width = 0.40\textwidth]{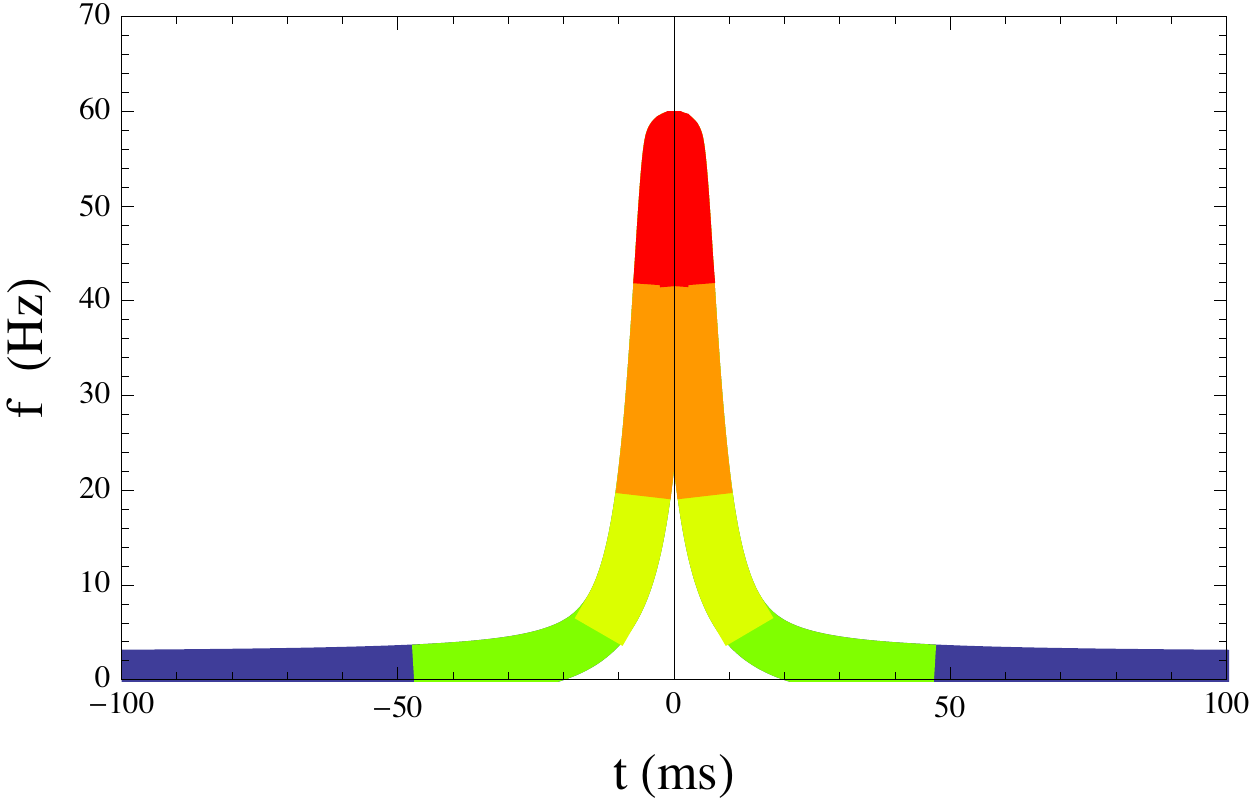}
\caption{The time dependence of the strain amplitude and frequency shift of GW in hyperbolic encounters, for the case $\beta=0.1$, $b=10^{-5} \textrm{AU}$ and $M=20~\Msun$. The different colors code different amplitudes, according to the top figure, and the peak frequency corresponds to  maximum GW emission.}
\vspace*{-2mm}
\label{fig:tfb}
\end{figure}

We thus consider a hyperbolic encounter between a BH of mass $m_2$ with asymptotic velocity $v_0$ against another BH of mass $m_1=q\,m_2\geq m_2$, see Fig.~\ref{fig:hyperbolic}. The total mass of the system is then given by $M=(1+q)\,m_2$ and the reduced mass is $\mu = q\,M/(1+q)^2$. For an impact parameter $b$, the eccentricity of the hyperbolic orbit is given by $e \equiv \sqrt{1+b^2v_0^4/G^2M^2}$~\cite{DeVittori:2012da}, and the orbital trajectory is characterized in polar coordinates by $r(\varphi) = a\,(e^2-1)/(1+e\cos(\varphi - \varphi_0))$, where $b=a\sqrt{e^2-1}$, and $\varphi_0 = {\rm arccos}\left(-1/e\right)$. Then, the strain amplitude and power emitted in GW are given by
\ba\label{strain}
h_c &=& \frac{2G}{R\,c^4}\langle\ddot Q_{ij}\ddot Q^{ij}\rangle^{1/2}_{i,j=1,2}
= \frac{2G\mu \,v_0^2}{R\,c^4}\,g(\varphi,\,e)\,, \\ \label{power}
\frac{dE}{dt} &=& - \frac{G}{45\,c^5}\langle \stackrel{\cdots}{Q}_{ij}
\stackrel{\cdots}{Q}{}^{\!ij}\rangle =
- \frac{32G\mu^2v_0^6}{45\,c^5\,b^2}\,p(\varphi,\,e) \,,
\ea
where $Q_{ij}$ is the reduced quadrupole moment of the BH encounter, and $p(\varphi,\,e)$ and $g(\varphi,\,e)$ are complicated bell-shaped functions of the angle $\varphi$, centered at $\varphi_0$. Here $R$ corresponds to the distance from us, which in practice is the luminosity distance $d_L(z)$ of the event. It can then be shown that the maximum values of the power and strain amplitude only depend on the eccentricity of the orbit, $p_{\rm max}(e) = 9(e+1)^2/(e-1)^4$ and $g_{\rm max}(e) = 2\sqrt{18(e+1)+5e^2}/(e-1)$. The time dependence of these functions is given by, with $u=\varphi-\varphi_0$,
\be\label{time}
\hspace*{-1mm}\frac{v_0\,t}{b} = \frac{e\sin u}{1+e\,\cos u}-\frac{2}{\sqrt{e^2-1}}{\rm tanh}^{-1}\!\!\left[\sqrt{\frac{e-1}{e+1}}\tan\frac{u}{2}\right]
\ee
from which we can estimate the characteristic time-scale of the event, $2\,t_{1/2}(e)$, corresponding to the full width at half maximum of the emission. In Fig.~\ref{fig:tfb} we show the time dependence of the frequency shift and strain amplitude of GW in hyperbolic encounters, for the case $\beta=v_0/c=0.1$, $b=10^{-5} \textrm{AU}$ and $M=20~\Msun$. The colored regions correspond to different GW amplitudes. As can be seen, the shape of the emission in the time-frequency domain is exactly as expected for a GW burst, similar to those ``tear-drop" bursts already observed in AdvLIGO~\cite{Powell:2016rkl}.

It is useful to express the maximum strain amplitude (\ref{strain}) and power (\ref{power}) in terms of physical quantities
\ba\label{hmax}
h_c^{\rm max}(e) &=& 3.24\times10^{-23}\, \frac{R_s({\rm km})}{d_L({\rm Gpc})}\,
\frac{q\,\beta^2\,g_{\rm max}(e)}{(1+q)^2}\,,\\ \label{Pmax}
P_{\rm max}(e) &=& 5.96\times10^{26}\, {\cal L}_\odot\
\frac{q^2\,\beta^{10}}{(1+q)^4}\,\frac{(e+1)}{(e-1)^5} \,,
\ea
where the solar luminosity is ${\cal L}_\odot = 3.9\times10^{33}$\,erg/s, and $R_s=2GM/c^2 = 3\,{\rm km}\,M/\Msun$ is the Schwarzschild radius. The maximum frequency of the GW emission corresponds to $f_{\rm peak} = 0.32\,{\rm mHz}\,(e+1)/(e-1)\cdot\beta/b$(AU), with the impact parameter in astronomical units. The product $f_{\rm peak}\cdot t_{1/2}(e)$ is a pure number that only depends on the eccentricity of the hyperbolic orbit. Since this can always be measured by the detector, we can estimate from here the parameter $e$ and, substituting in (\ref{hmax}) and (\ref{Pmax}), determine $q$ and $\beta$ for a given distance to the source. Note that the eccentricity provides a direct connection between the orbital parameters,
\be
e^2=1+\left(\frac{b}{10^{-5}\, {\rm AU}}\right)^2
\left(\frac{\beta}{0.1}\right)^4\left(\frac{10\,\Msun}{M}\right)^2\,.
\ee
For concreteness, we will now consider some typical examples in both the AdvLIGO and LISA range.

\begin{figure}[!h]
\centering
\includegraphics[width = 0.48\textwidth]{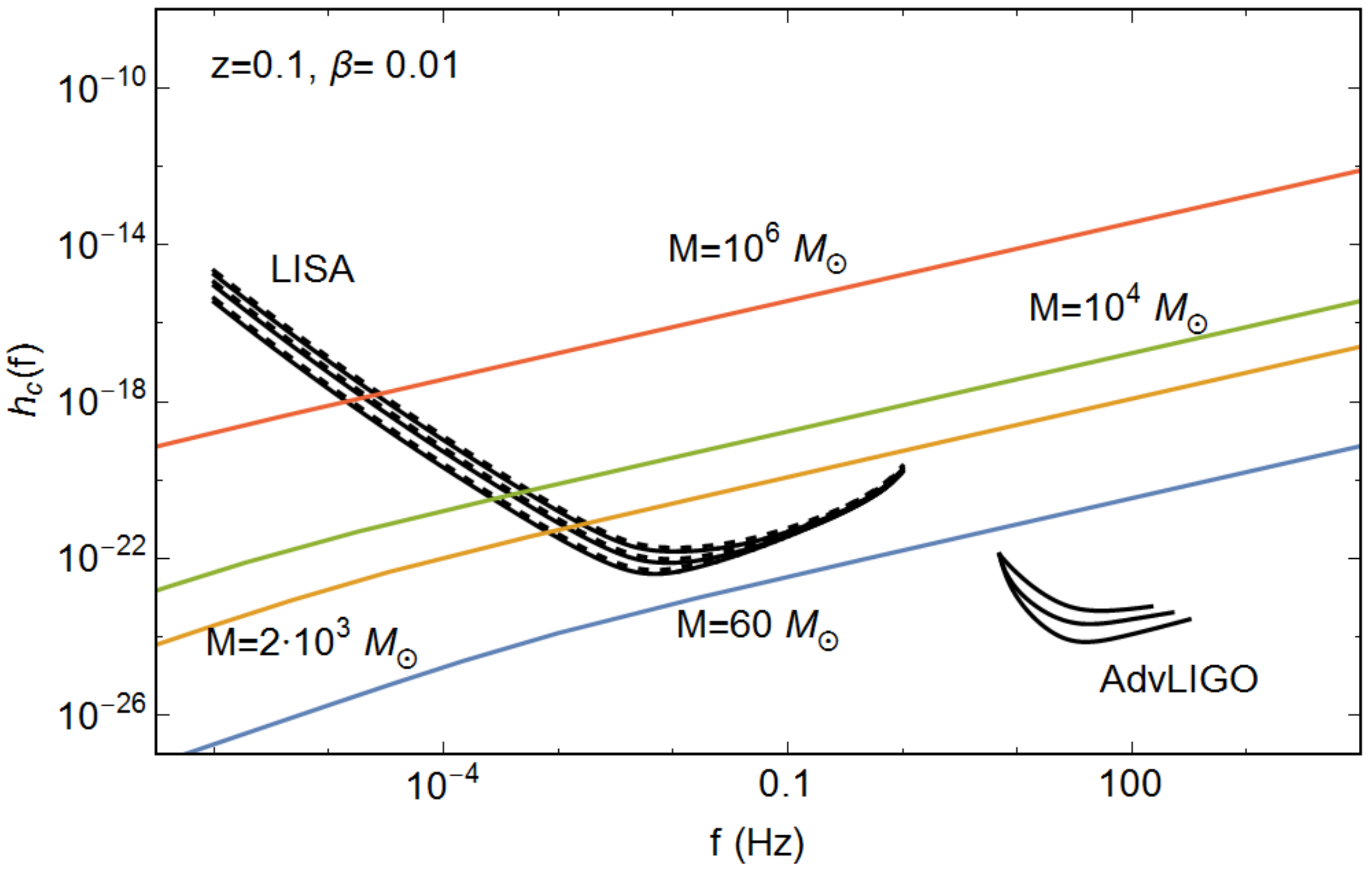}
\caption{The AdvLIGO and LISA sensitivity curves, together with the locus of peak frequencies, as a function of impact parameter,
$\log_{10}(b/AU)\in[-6,3]$, for different PBH total masses, $60 - 10^6\,\Msun$, and a redshift to the source of $z=0.1$.}
\vspace*{-1mm}
\label{fig:sensitivity}
\end{figure}

\subsection{GW bursts in AdvLIGO}

Let us consider first the hyperbolic encounter of two BH of $30~\Msun$ each, moving at a relative speed $\beta = 0.1$, with impact parameter $b=2\times10^{-5}$ AU. The eccentricity of the hyperbolic orbit is $e=1.054$ and the maximum power emitted in GW is given by $P_{\rm max} = 6.44\times10^{55}\, {\rm erg/s}$, or $1.65\times10^{22}$ times larger than the solar luminosity in EM waves. We can also compute the maximum stress amplitude that such an event would induce on a laser interferometer on Earth, at a distance of $d_L=1$ Gpc, $h_c^{\rm max} = 3.55\times10^{-21}$, while the duration of the event can be easily computed from Eq.~(\ref{time}), and is given by $\Delta t = 2\,t_{1/2}(e) \simeq 7.6$\,ms. The corresponding maximum frequency is $f_{\rm peak} \simeq 60.4$\,Hz, which lies perfectly within the AdvLIGO sensitivity band.

\subsection{GW bursts in LISA}

Second, let us consider here a close encounter of an IMBH of mass $m_2 = 10^3\,\Msun$ and a SMBH of mass $m_1 = 10^6\,\Msun$, as expected at the centers of galaxies. The impact parameter $b=1$\,AU and velocity $v_0 = 0.05\,c$ gives an eccentricity parameter of $e=1.031$ and a maxi\-mum power emitted $P_{\rm max} = 1.66\times10^{49}$\,erg/s, which is $4.26\times10^{15}$ times larger than solar luminosity. The maxi\-mum stress amplitude, at a distance $d_L=1$ Gpc, is in this case $h_c^{\rm max} = 1.02\times10^{-19}$. The duration of the event is given by $\Delta t \simeq 440$\,s, while the corresponding peak frequency is $f_{\rm peak} = 1.05$\,mHz, which lies nicely within the sensitivity band of LISA~\cite{Bartolo:2016ami}.

Alternatively, we can consider a hyperbolic encounter between two SMBH of equal masses $m_1=m_2=10^6\,\Msun$, with an impact parameter $b=10$ AU and relative velocity $v_0=0.015\,c$, possibly occurring during galaxy collisions at low redshift. The eccentricity is low, $e=1.01$, and the stress amplitude is huge $h_c^{\rm max} = 2.22\times10^{-17}$, easily detectable by LISA, with a duration of one day, and a peak power $P_{\rm max} \simeq 1.68\times10^{52}$ erg/s, at $f_{\rm peak} = 1.51\times10^{-4}$ Hz, right in the middle of LISA sensitivity. Such an event would be clearly distinguishable.

\begin{figure}[!h]
\centering
\includegraphics[width = 0.48\textwidth]{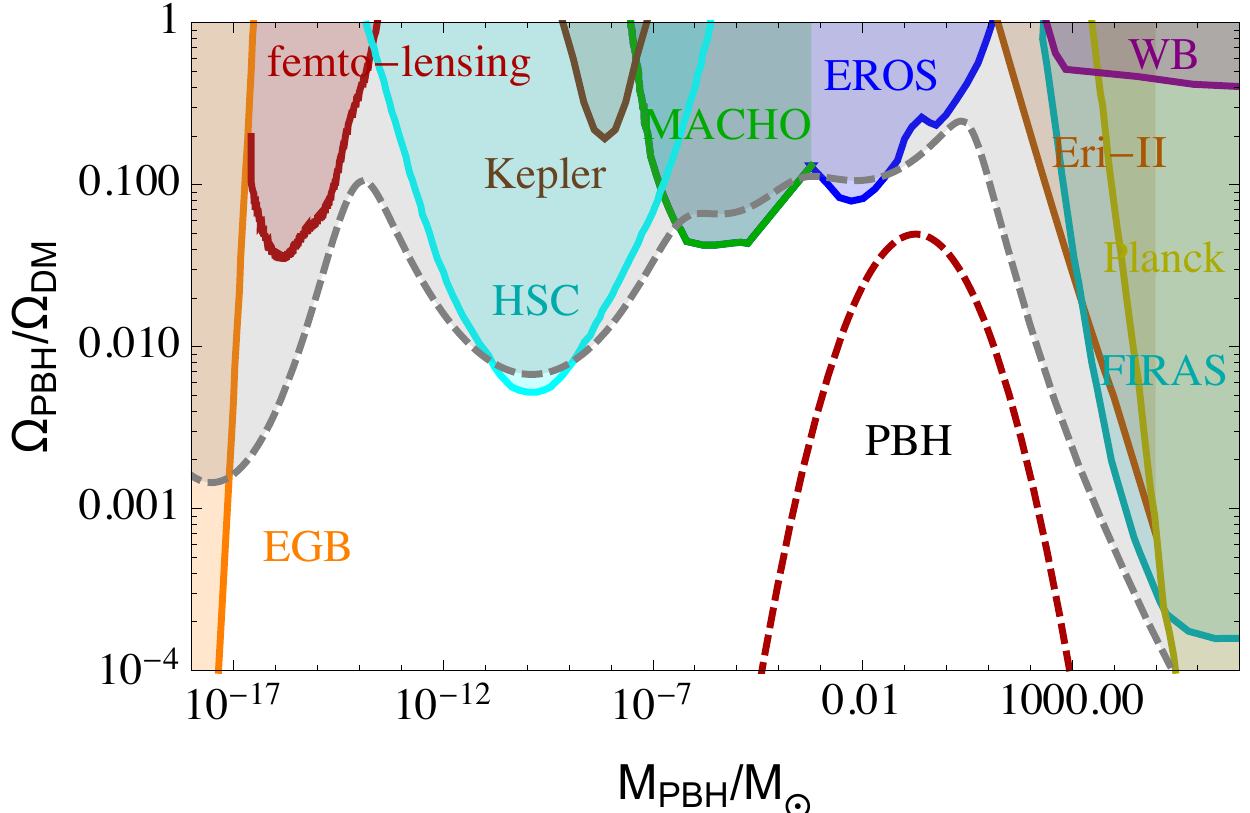}
\caption{Observational constraints on PBH from a plethora of experiments (for a review see~\cite{Carr:2016drx}). We have taken into account the fact that our mass distribution is wide, which changes the constraints (gray line), for details see Ref.~\cite{Carr:2017jsz}.}
\vspace*{-1mm}
\label{fig:constraints}
\end{figure}

\begin{figure*}[!t]
\includegraphics[width = 0.329\textwidth]{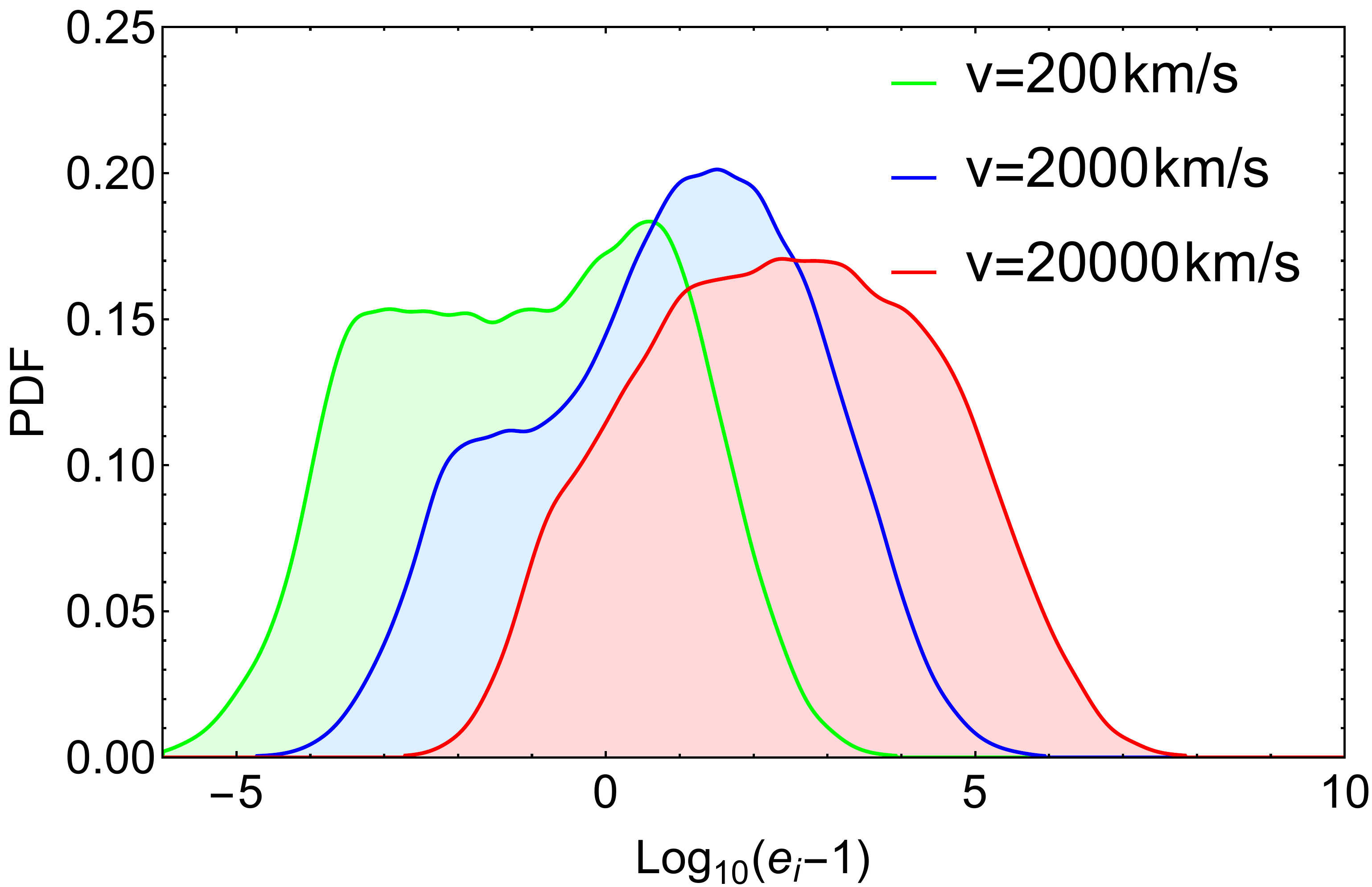}
\includegraphics[width = 0.328\textwidth]{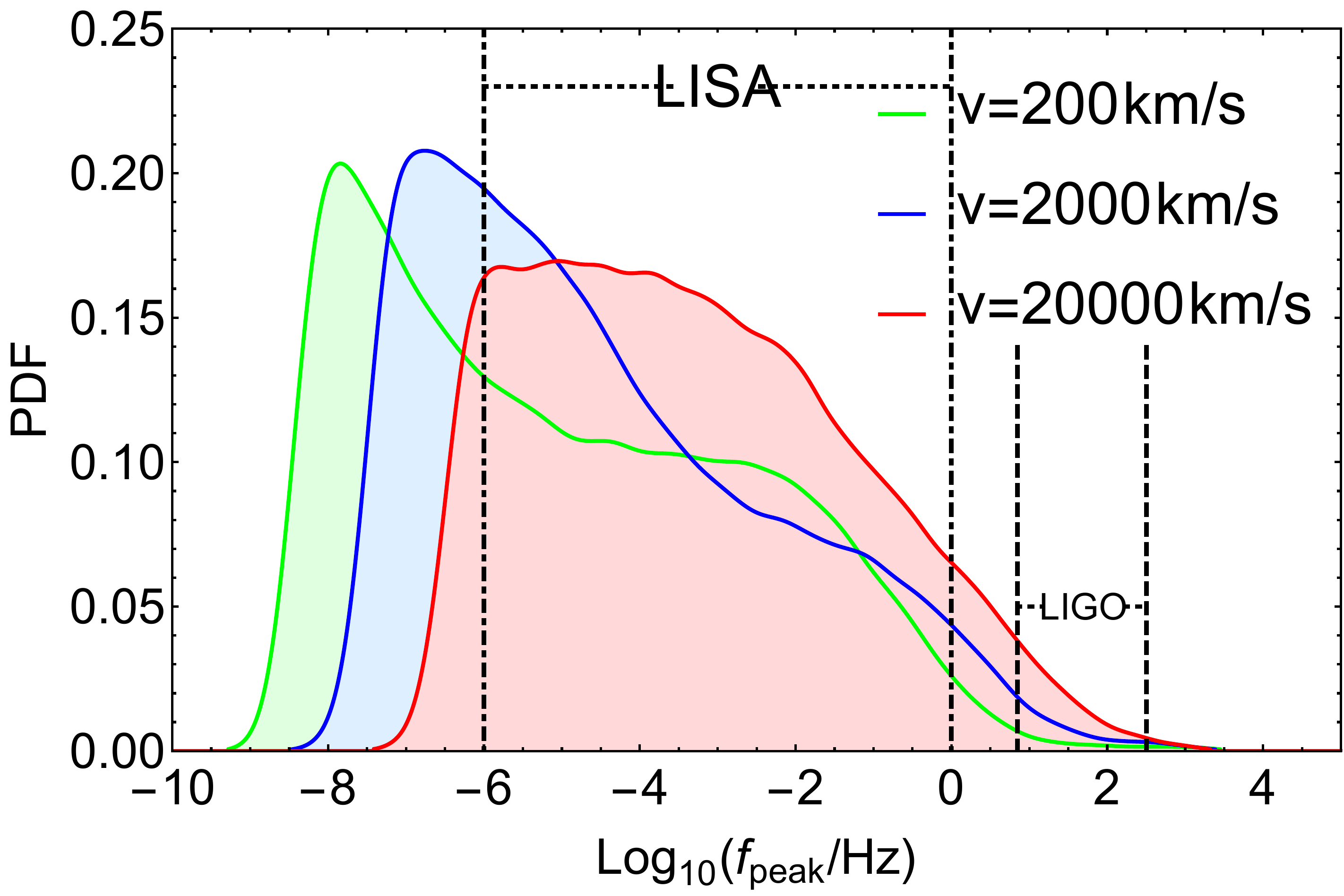}
\includegraphics[width = 0.329\textwidth]{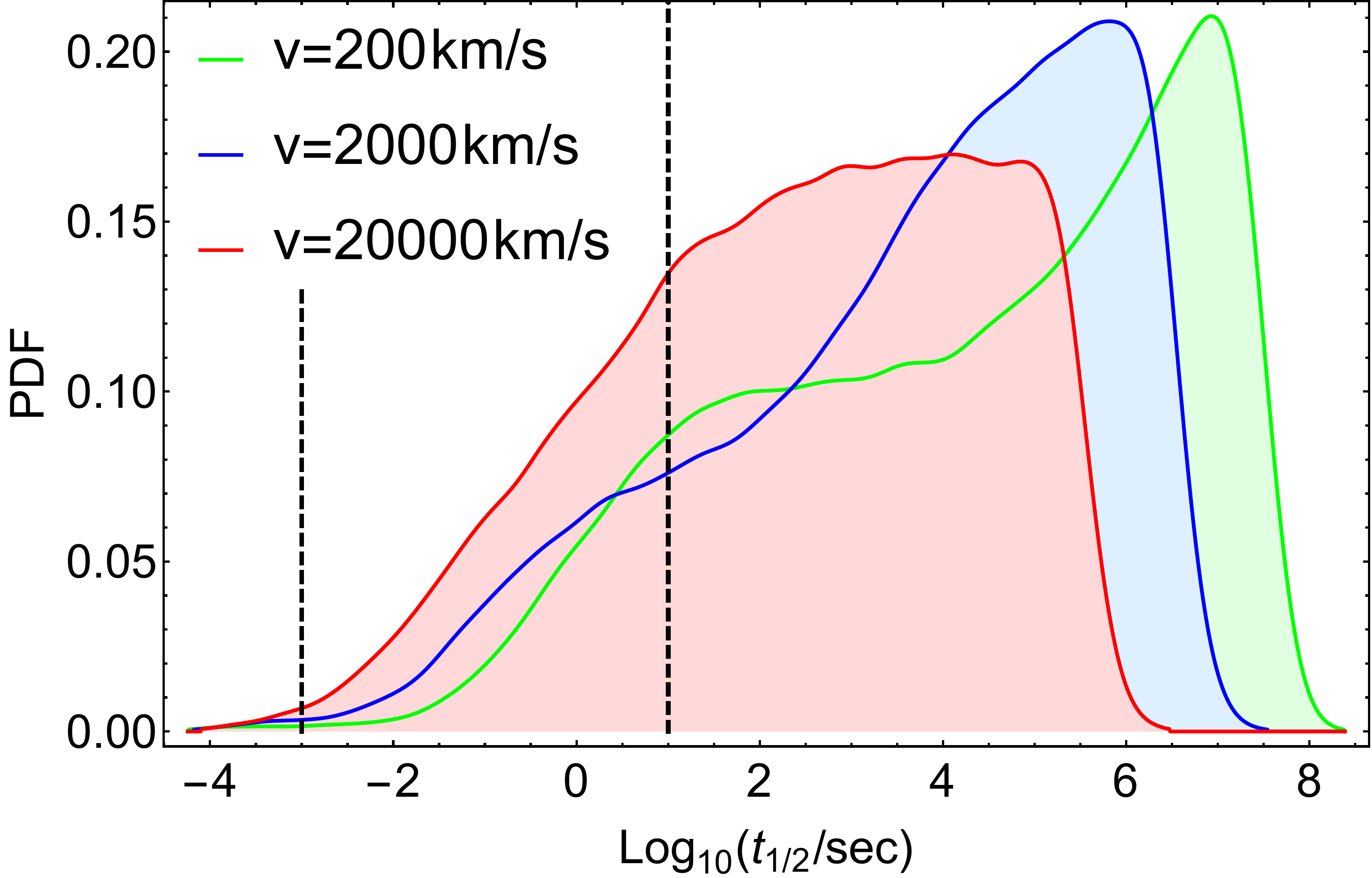}
\caption{The population rates for a log-normal distribution of PBH with parameters of a mean mass equal to $60\,\Msun$, deviation  $\sigma=2.4$ and a flat prior in $\log_{10}{b}$ for $b\in[b_{\rm min},\,100\,\textrm{AU}]$, where $b_{\rm min}$ is the minimum value that satisfies the constraints discussed in the text. The panels show the distributions for various velocities $v$ of the PBH, as indicated by the colored lines, for the eccentricity $e$ (left), the peak frequency of GW emission $f_{\rm peak}$ (center) and the characteristic timescale $t_{1/2}$ (right). The two vertical dashed lines (right) correspond to time scales of a few milliseconds to a few tens of seconds, as expected for bursts.}
\label{fig:rates}
\end{figure*}

\section{Estimating event rates: LIGO \& LISA}

We explore the parameter space accessible to Adv\-LIGO and LISA in terms of the impact parameter $b$ and the velocity $\beta$, with the constraint that the peak frequency $f_{\rm peak}$ is within the domain of the detectors, and the peak strain $h_c^{\rm max}$ is above the sensitivity curves, see Fig.~\ref{fig:sensitivity}. We estimate, following Ref.~\cite{Clesse:2016ajp}, the probability of finding an event in LIGO and LISA by simulating the BH encounters with impact parameters drawn from a flat prior in $\log_{10}b$, with $b\in[10^{-6},10^2]$\,AU, and masses given by a normalized lognormal distribution
\be
P(M) = \frac{f_{\rm PBH}}{M\sqrt{2\pi}\,\sigma}\,\exp\left[-\frac{\ln^2(M/\bar\mu)}{2\sigma^2}\right]
\ee
with $\bar\mu = 60\,\Msun$ and $\sigma=2.4$. Since the mass distribution is very wide, the PBH constraints are modified w.r.t. the monochromatic ones of Ref.~\cite{Carr:2016drx}, see e.g. Ref.~\cite{Carr:2017jsz} and Fig.~\ref{fig:constraints}. We assume through out that our mass distribution gives a total PBH/DM fraction $f_{\rm PBH} = 1$.

We impose constraints on the impact parameter $b$ so that the BHs remain unbound, and the impact parameter is larger than the Schwarzschild radius $R_s = 2GM/c^2$ and we also consider relative speeds $v_0<0.1\,c$, in order to remain in the non-relativistic regime~\cite{JGBSN:2017}. The results of these simulations are the distribution functions for the eccentricity $e$, the peak frequency of the emission $f_{\rm peak}$ and the characteristic timescale $t_{1/2}$, as shown in Fig.~\ref{fig:rates}. We estimate that a large fraction of the PBH population is expected to have peak frequencies within the LISA or AdvLIGO ranges, with relatively large eccentricities $(e\sim1)$, and typical time-scales above a few milliseconds.

\begin{figure*}[!t]
\includegraphics[width = 0.45\textwidth]{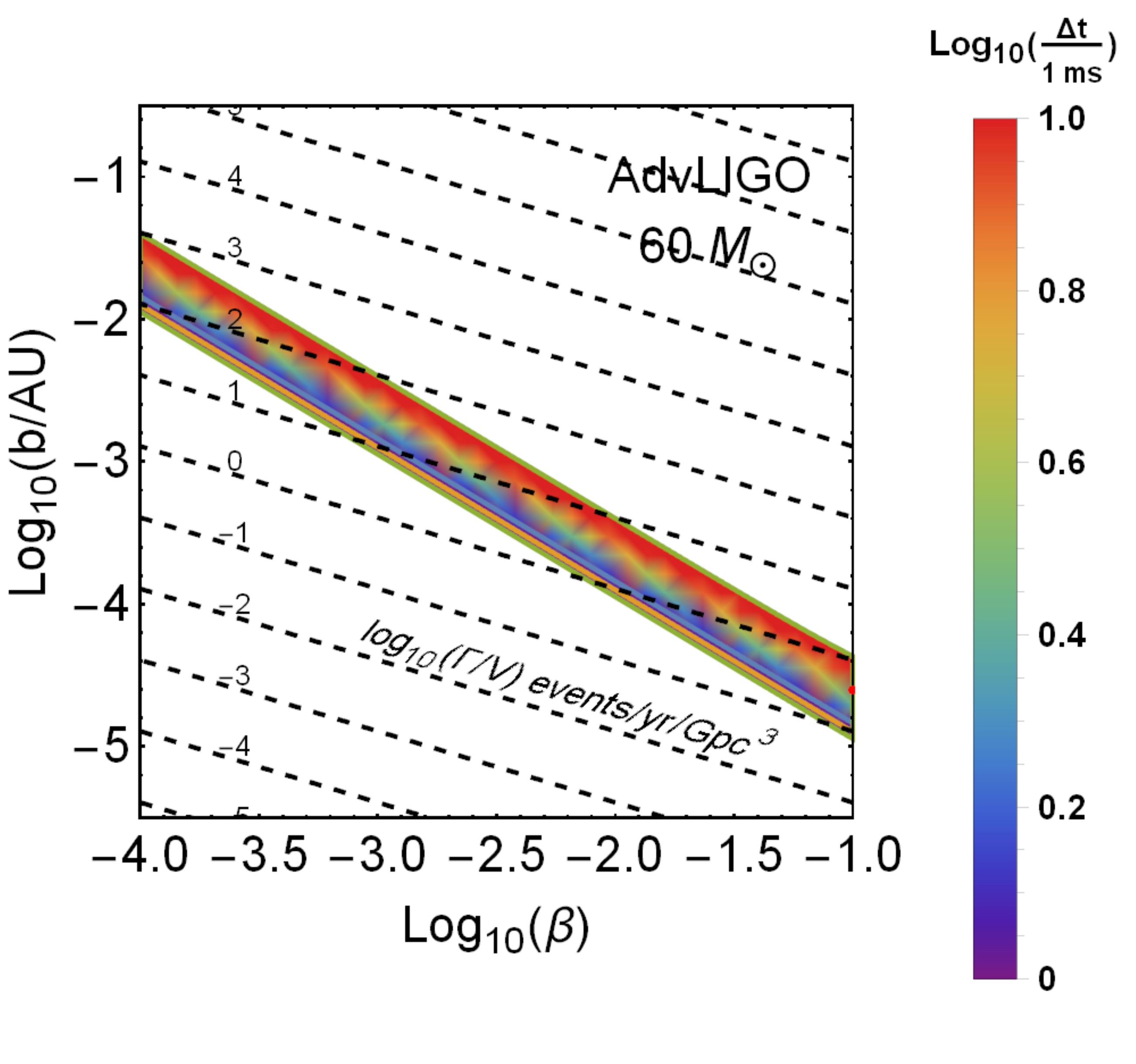}\hspace{5mm}
\includegraphics[width = 0.45\textwidth]{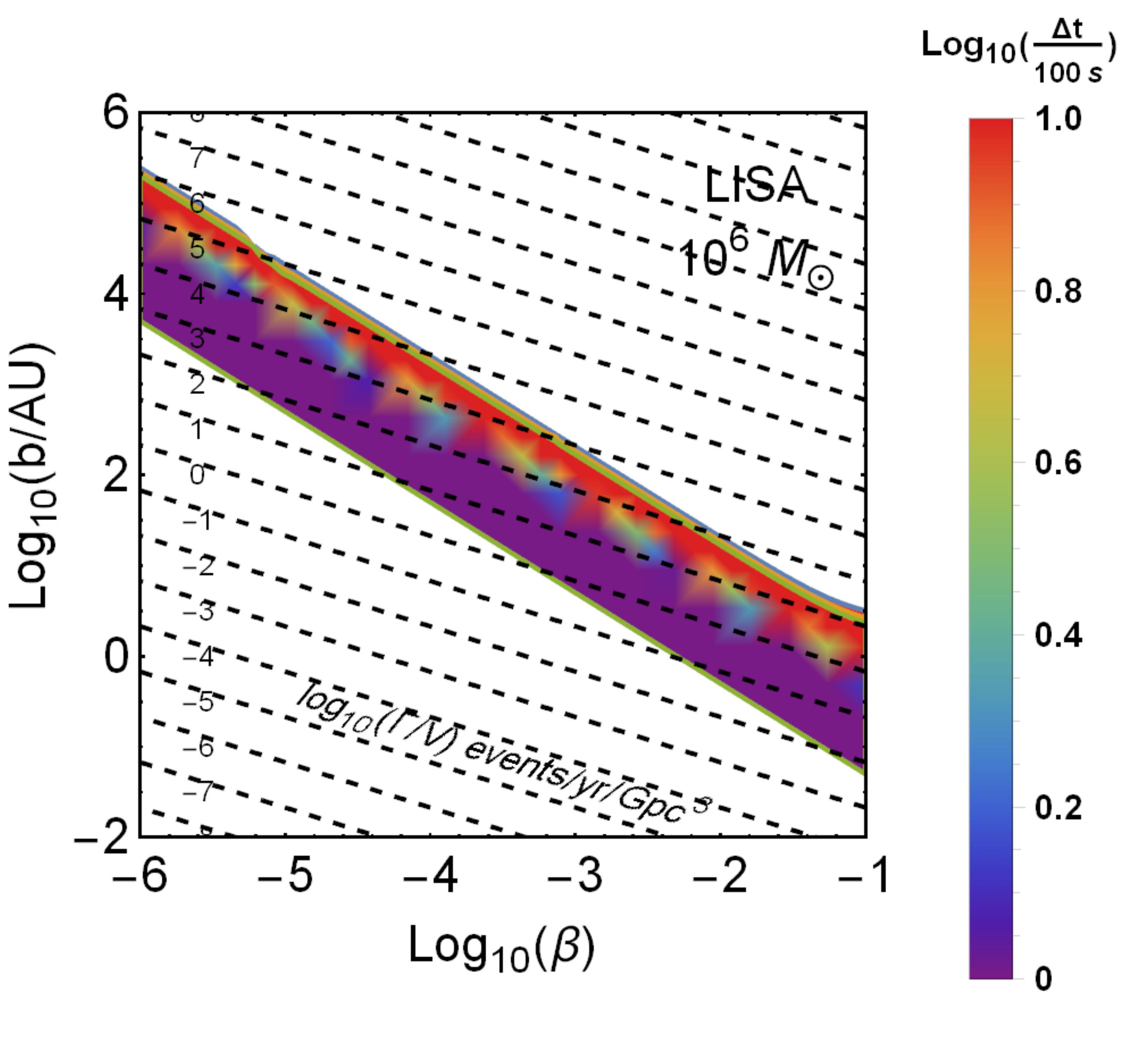}
\vspace{-3mm}\caption{The parameter space accessible to AdvLIGO and LISA sensitivities, in terms of the relative velocity $\beta$ and the impact parameter $b/\textrm{AU}$, using the distributions of Fig.~\ref{fig:rates}. In all cases we assume the BH pair to be at a redshift of $z=0.1$. We show the parameter space for PBH encounters of $60\Msun$ (left) for AdvLIGO, and $10^6 \Msun$ (right) for LISA. The coloring scheme corresponds to the characteristic timescale of the event, while the dashed lines show the expected event rate in units of events/yr/Gpc$^3$, i.e. $\log_{10}(\Gamma/V)$. The band width is related to the sensitivity of the detectors, see Fig.~\ref{fig:sensitivity}.}
\label{fig:paramspace}
\end{figure*}

The estimated rate of events per unit volume that Adv\-LIGO and LISA can detect within their sensitivity range can be obtained as~\cite{Capozziello:2008ra,JGBSN:2017}
\ba
\Gamma/V&\simeq&n^2 v_0~\sigma \nn
\ea
where $n=\rho_{\rm PBH}/M$ is the number density of PBHs, $v_0$ the relative velocity and $\sigma=\pi b^2$ the cross-section of the encounter. Using that $\rho_{\rm PBH}=\Omega_{\rm DM}~f_{\rm PBH}~\delta_{\rm loc}~\rho_{cr}$, that $h=0.7$ and $\Omega_{\rm DM}=0.25$, we get that
\ba
\Gamma/V &=& \frac{9}{64\pi}\frac{H_0^4}{v_0^3}\,f^2_{\rm PBH}\,\delta^2_{\rm loc}\,\Omega^2_{\rm DM}\,(e^2-1) \nn \\
&=& 25.4\ {\rm yr}^{-1}{\rm Gpc}^{-3}\left(\frac{\delta_{\rm loc}}{10^8}\right)^2
\frac{e^2-1}{\beta^3}
\ea
where the parameter $\delta_{\rm loc}$ is the local density contrast of the PBH cluster~\cite{Clesse:2016vqa,Clesse:2016ajp}, which can give a significant enhancement to the rate of GW bursts events, compared with previous studies~\cite{DeVittori:2012da}. Furthermore, the rate could be also enhanced for $\beta\ll1$.

In Fig.~\ref{fig:paramspace}, the dashed lines show the expected event rate in units of events/yr/Gpc$^3$. For a wide range of impact parameters and relative velocities, the expectation is that the interferometers will detect a few events per year for hyperbolic encounters of PBH of tens of solar masses in dense clusters, and even larger for the scattering of SMBH in the center of colliding galaxies. Furthermore, in the case of AdvLIGO we find that the characteristic timescale of the events is of the order of milliseconds, while for LISA is a few hours.

\section{Conclusions}

We have found that hyperbolic encounters of PBH can produce GW bursts with frequencies, strain amplitudes and characteristic time durations within the sensitivity of the AdvLIGO and LISA interferometers, as summarized in Figs.~\ref{fig:sensitivity} and \ref{fig:paramspace}. These events would have unique signatures, very different from the usual inspiralling stellar BH, and would provide strong evidence in favor of the clustered PBH scenario. The spectral power density, peak frequency, maximum amplitude and event duration give us direct information about the orbital parameters of the scattering PBH, while the event rates tell us about their spatial distribution. We have found event rates, for both AdvLIGO and LISA, of the order of a few events/year/Gpc$^3$, thus making them readily available for observation with current and future detectors. Moreover, since all the calculations can be done for small enough velocities and large impact parameters, we do not need to rely on sophisticated numerical relativity codes to provide templates for a proper comparison with the already available data from the AdvLIGO Collaboration.

\section*{Acknowledgements}

The authors acknowledge support from the Research Project FPA2015-68048-03-3P [MINECO-FEDER] and the Centro de Excelencia Severo Ochoa Program SEV-2012-0597. JGB acknowledges support from the Salvador de Madariaga Program, Ref. PRX17/00056. S.N. also acknowledges support by the Ram\'on y Cajal Program through Grant No. RYC-2014-15843.


\end{document}